# First-Principles Studies of Second-Order Nonlinear Optical Properties of Organic-Inorganic Hybrid Halide Perovskites

Wenshen Song[1], Guang-Yu Guo[2,3*], Su Huang[4], Lan Yang[5], and Li Yang[1,6*]

[1]*Department of Physics, Washington University in St. Louis, St. Louis, Missouri 63130, USA*

[2]*Department of Physics and Center for Theoretical Physics, National Taiwan University, Taipei 10617, Taiwan.*

[3]*Physics Division, National Center for Theoretical Sciences, Hsinchu 30013, Taiwan*

[4]*Aerosol and Air Quality Research Laboratory, Department of Energy, Environmental, and Chemical Engineering, Washington University in St. Louis, St. Louis, MO 63130, United States*

[5]*Department of Electrical and Systems Engineering, Washington University, St. Louis, MO 63130, USA*

[6]*Institute of Materials Science and Engineering, Washington University in St. Louis, St. Louis, Missouri 63130, USA*

## Abstract

Organic-inorganic hybrid halide perovskites have ignited tremendous interests for photovoltaic applications. However, their nonlinear optical response has not been studied although many of these structures lack the centrosymmetry and exhibit ferroelectricity. In this work, we employ our developed large-scale parallel, first-principles simulation tool (ArchNLO) to explore second-order nonlinear optical properties of a typical family of organic-inorganic hybrid halide perovskites, $CH_3NH_3MX_3$ (M= Ge, Sn, Pb; X=Cl, Br, I). We find that these hybrid perovskites exhibit second harmonic generation and linear electro-optic effect. The nonlinear optical effects are strongly influenced by the types and positions of cations/anions and corresponding band gaps. Particularly, the distorted cubic phase, which is essentially triclinic, of $CH_3NH_3SnI_3$ shows significant second harmonic generation and electro-optic effect, which are comparable with those widely used materials, such as GaAs. These second-order optical properties of organic-inorganic hybrid halide perovskites and their low-temperature, solution-based fabrication pave the way for achieving and implementing nonlinear optical devices with low cost.

## I. Introduction

Nonlinear optical (NLO) materials play a critical role in modern electronics and photonics by providing means to alter the phase, frequency or amplitude of input electromagnetic waves. Such alternation can be realized through a variety of nonlinear processes, such as the linear electro-optic (LEO) effect, second/third/fourth/high harmonic generation (SHG/THG/FHG/HHG), and the Kerr effect, etc. [1] The group of traditional NLO materials are ternary inorganic oxides and their derivatives., e.g., lithium niobate and lithium tantalate. They have achieved widespread success due to their reliable performance, low optical loss, and good stability. However, the synthesis of these insoluble oxides requires the high temperature treatment, which hinders broader applications for flexible substrates and the integration into chip-scale nanophotonic devices. In contrast to their inorganic counterparts, organic NLO materials based on chromophores have been considered as a promising alternative due to their solution processability, faster response, and stronger NLO activities. Unfortunately, the low intrinsic stability and high optical loss severely limit their applications [2].

Nowadays, organic-inorganic hybrid halide perovskites (OHPs) have attracted tremendous interest in emerging photovoltaic (PV) technologies. Extensive theoretical exploration has been applied to studying their linear optical responses related to PV, including self-consistent density functional theory (DFT) approaches as well as quasiparticle methods [3–5]. On the other hand, due to the asymmetry of the organic unit, ferroelectricity can be realized in this family of materials naturally or by artificially tuning [6–9]. This broken inversion symmetry also ensures NLO properties [10], especially second harmonic generation (SHG) and linear electro-optic (LEO) effects [11–13]. Moreover, given the facts that such compounds can be fabricated from the liquid phase [14,15] and nearly unlimited choices of cations and anions [16,17], OHPs may exhibit unique advantages in fabricating cost and optimizations of NLO properties.

Calculating and predicting NLO properties of OHPs are challenging for available first-principles simulation tools [18–21]. Compared to those of linear optical responses, NLO calculations involve higher-order transitions, which are intrinsically much more expensive in simulation cost. To date most NLO studies focus on materials either with a small number of atoms per unit cell or low-dimension structures, which can reduce the number of empty bands or the k-point samplings in the reciprocal space. Unfortunately, OHPs contain a larger number of atoms with relatively low symmetries. Thus, studying their NLO properties needs fundamental coding developments for better parallelization performance and simulation efficiency.

In this work, we have developed a high-efficient, parallel code (ArchNLO) based on density functional theory (DFT) to calculate second-order NLO properties of materials. Using Fortran with the Message Passing Interface (MPI) and parallel computations on k-points, an excellent parallelization performance can be held for up to a few thousand processors. This development enables us to calculate large-scale systems with low symmetries. Our first-principles calculations systematically investigate an important family of OHPs, *i.e.*, $CH_3NH_3MX_3$ (M=Ge, Sn, Pb; X= I, Br, Cl). SHG and the LEO effect are observed in these structures, and they are strongly influenced by the types and positions of cations/anions and corresponding band gaps. Particularly, $CH_3NH_3SnI_3$ exhibits significant SHG and LEO effect, which are comparable with traditional NLO materials. We further analyze the mechanism for such large SHG and LEO coefficients and their relationships with electronic structures.

The remainder of this paper is organized as follows: In Sec. II, we introduce the atomic structures of OHPs and our computational approaches including formulas of linear and NLO calculations. In Sec. III, we present the simulation performance of ArchNLO. In Sec. IV, the electronic band structures of our studied OHPs are presented. In sec. V, the linear optical response of OHPs is presented. In Sec. VI, we show SHG of OHPs and discuss the details. In Sec. VII, we present the LEO effect and discuss the mechanism behind. In Sec. VIII, we summarize our studies and conclusion.

## II. Atomic Structures and Computational Details

The first-principles calculations to obtain Kohn-Sham wavefunctions and band structures are based on DFT within the generalized gradient approximation (GGA) by using the Perdew-Burke-Ernzerhof (GGA-PBE) functional [22], implemented in the Vienna ab initio simulation package (VASP) [23,24]. The van der Waals (vdW) interaction is included through the DFT-D3 method with the Becke-Jonson damping [25,26]. The plane-wave energy cutoff is set to be 450 eV to guarantee the convergence of electronic and optical calculations [27,28]. We use a hybrid functional with a fraction of alpha = 0.45 exact exchange with a range-separation identical to the Heyd-Scuseria-Ernzerhof (HSE06) hybrid functional [29,30] to correct underestimated DFT band gaps. Spin-orbit coupling (SOC) is included as well. We use a 4 × 4 × 4 k-grid sampling in the reciprocal space to solve the Kohn-Sham equation.

We start from the prototype cubic phase of $CH_3NH_3MX_3$ (M=Ge, Sn, Pb; X= I, Br, Cl) [31,32], which are usually observed at room temperature, making them of application interests. The atomic structure is illustrated in Figure 1a. In a unit cell, the metal M atom sits on vertices, and the halide X atom sits on the middle position of each edge, forming $MX_6$ octahedra cages around the centered organic unit $CH_3NH_3$.

We fully relax structures until DFT/PBE-calculated forces are smaller than 0.01 eV/Å, and the relaxed structural parameters are summarized in Table 1. Because we do not apply any symmetry constrain during relaxation, our cubic structure is distorted by the anisotropic $CH_3NH_3$ molecule that ends up along the [101] direction. The similar distortions are also observed in previous simulations of cubic-phase $CH_3NH_3PbI_3$. **(**$a = 6.28$ Å$, b = 6.38$ Å$, c = 6.34$ Å$, \alpha = 91.7°, \beta = 90.0°, \gamma = 90.0°$) [33]. Because all symmetries are broken in simulations, our relaxed structures are essentially the triclinic phase despite the distortions are small (1~2%). The corresponding first Brillouin zone (BZ) and high-symmetry points are presented in Figure 1b. Our DFT-calculated structures show good agreements with previous works [4,31,33,34]. For example, our calculated lattice parameters of the $CH_3NH_3PbI_3$ are $a = 6.31$ Å$, b = 6.23$ Å$, c = 6.38$ Å$, \alpha = 90.0°, \beta = 89.2°, \gamma = 90.0°$, which are close to measurements [34] ($a = 6.33$ Å$, \alpha = \beta = \gamma = 90°$).

*Calculation of Optical Properties:* The linear optical dielectric function and NLO susceptibility are calculated based on the linear response formalism within the independent-particle approximation (IPA) [35–38]. The formulas are adapted from the reference [35] with slight modifications [39]. Because of the large unit cell of OHPs, the GW method [40], an established *ab initio* approach for obtaining reliable quasiparticle band gaps, is formidable for our current simulation capability. Thus, we employ the hybrid functional and the scissor approximations to correct DFT band gaps [21,35,39,41], which have been widely used in previous works [4,27,33]. More sophisticated approaches, such as the "layer-by-layer" analysis [42], will be implemented in the future for improving the calculation of optical responses.

Excitonic effects are not included in this work, due to the extremely high computational expense, especially for OHPs with a large number of electrons per unit cell. Moreover, electron-hole interactions are usually more significant in reduced-dimensional structures or large-gap insulators [43]. For example, in bulk GaAs, we find that the independent-particle picture with a corrected band gap can capture the main features of its SHG spectrum, such as peak positions and the low-frequency amplitude (see Section I of the Supplementary Material [44]).

The linear susceptibility is given by [35]

$$\chi_I^{ab}(-\omega;\omega) = \frac{e^2}{\hbar}\sum_{nmk} f_{nm}\frac{r_{nm}^a(\boldsymbol{k})r_{mn}^b(\boldsymbol{k})}{\omega_{mn}(\boldsymbol{k})-\omega}, \quad \text{Eq. 1}$$

in which $n, m$ are bands indices, and $f_{nm}$ is the difference of Fermi-Dirac distributions between two states. The momentum matrix element, $p_{ij}^a = \langle \boldsymbol{k}j|\hat{p}_a|\boldsymbol{k}i\rangle$, is the transition between two states $i$ and $j$ at the $\boldsymbol{k}$ point, and the position matrix element is defined by $r_{nm}^a(k) = \frac{p_{nm}^a(k)}{im\omega_{nm}}$ if $n \neq m$ or 0 if $n = m$.

The SHG susceptibility is calculated by [35,36]

$$\chi_{abc}^{(2)}(-2\omega;\omega,\omega) = \chi_{II}^{abc}(-2\omega;\omega,\omega) + \eta_{II}^{abc}(-2\omega;\omega,\omega) + \sigma_{II}^{abc}(-2\omega;\omega,\omega), \quad \text{Eq. 2}$$

in which the first term $\chi_{II}^{abc}(-2\omega;\omega,\omega)$ is from the interband transitions at the same crystal momentum $\boldsymbol{k}$; the second term $\eta_{II}^{abc}(-2\omega;\omega,\omega)$ is the intraband contribution from the modulation of the linear susceptibility due to intraband motions of electrons; the third term $\sigma_{II}^{abc}(-2\omega;\omega,\omega)$ is the modification contribution, representing the modification by the polarization energy associated with interband motions. The detailed expressions are in the following [35,36].

The interband term is:

$$\chi_{II}^{abc}(-2\omega;\omega,\omega) = \frac{e^3}{\hbar^2}\sum_{nml}\int\frac{d\boldsymbol{k}}{8\pi^3}\frac{r_{nm}^a\{r_{ml}^b r_{ln}^c\}}{\omega_{ln}-\omega_{ml}}\left\{\frac{f_{ml}}{\omega_{ml}-\omega}+\frac{f_{ln}}{\omega_{ln}-\omega}+\frac{2f_{nm}}{\omega_{mn}-2\omega}\right\}. \quad \text{Eq. 3}$$

The intraband term is:

$$\eta_{II}^{abc}(-2\omega;\omega,\omega) = \frac{e^3}{\hbar^2}\int\frac{d\boldsymbol{k}}{8\pi^3}\left\{\sum_{nml}\omega_{mn}r_{nm}^a\left\{r_{ml}^b r_{ln}^c\right\}\left\{\frac{f_{nl}}{\omega_{ln}^2(\omega_{ln}-\omega)}+\frac{f_{lm}}{\omega_{ml}^2(\omega_{ml}-\omega)}\right\} - 8i\sum_{nm}\frac{f_{nm}r_{nm}^a}{\omega_{mn}^2(\omega_{mn}-2\omega)}\{\Delta_{mn}^b r_{mn}^c\} - \frac{2\sum_{nml}f_{nm}r_{nm}^a\left\{r_{ml}^b r_{ln}^c\right\}(\omega_{ln}-\omega_{ml})}{\omega_{mn}^2(\omega_{mn}-2\omega)}\right\}. \quad \text{Eq. 4}$$

in which $\Delta_{mn}^b = \frac{P_{mm}^b(k)-P_{nn}^b(k)}{m_e}$. ($m_e$ is the electron mass)

Finally, the modification term is:

$$\sigma_{II}^{abc}(-2\omega;\omega,\omega)=\frac{e^3}{2\hbar^2}\int\frac{d\boldsymbol{k}}{8\pi^3}\left\{\sum_{nml}\frac{f_{nm}}{\omega_{mn}^2(\omega_{mn}-\omega)}\left[\omega_{nl}r_{lm}^a\{r_{mn}^b r_{nl}^c\}-\omega_{lm}r_{nl}^a\{r_{lm}^b r_{mn}^c\}\right]+i\sum_{nm}\frac{f_{nm}r_{nm}^a\{\Delta_{mn}^b r_{mn}^c\}}{\omega_{mn}^2(\omega_{mn}-\omega)}\right\}. \quad \text{Eq. 5}$$

The LEO coefficient $r_{abc}(\omega)$ is connected to the LEO susceptibility $\chi_{abc}^{(2)}(-\omega,\omega,0)$ at the zero frequency limit as [21,35]:

$$r_{abc}(\omega)=-\frac{2}{\bar{n}^4(0)}\lim_{\omega\to 0}\chi_{abc}^{(2)}(-\omega;\omega,0)=-\frac{2}{\bar{n}^4(0)}\lim_{\omega\to 0}\chi_{abc}^{(2)}(-2\omega;\omega,\omega), \quad \text{Eq. 6}$$

where $\bar{n}^2(0)$ equals the geometric mean dielectric constant $\bar{\varepsilon}(0)=[\varepsilon_x(0)\varepsilon_y(0)\varepsilon_y(0)]^{1/3}$. Within the low-frequency range below the band gap, $\chi_{abc}^{(2)}(-2\omega,\omega,\omega)$ and $n(\omega)$ are nearly constant, so that the LEO coefficient $r_{abc}(\omega)\approx r_{abc}(0)$ [21,45].

We have to emphasize that, to obtain converged NLO properties, the *k*-point sampling needs to be much denser than that in self-consistent DFT band-structure calculations. For our OHPs, a k-point sampling of 32 × 32 × 32 with all symmetries broken and 120 total electronic bands (50 valence bands and 70 conduction bands) are included to guarantee the convergence of NLO calculations (see details in Section II of the Supplementary Material [44]).

## III. ArchNLO Performance

*Parallelism and scalability:* The ArchNLO package can accept the outputs from VASP, which include the momentum matrix elements, Kohn-Sham eigenvalues, and electronic occupation numbers. Then it separates data by k-points and implements parallel reading and computes in each node based on Fortran MPI. Our parallelization straightforwardly distributes k-points to each computing node independently, making large-scale calculations scalable. Beyond parallelization, we have substantially optimized the I/O communication by using the stream I/O to speed up the reading and writing processes. This optimization reduces the time of I/O processes by several orders of magnitude for large systems.

In parallel computations, scalability is a vital criterion to evaluate the simulation performance. It is a measure of a parallel system's capacity to increase speed in proportion to the number of processors [46]. An example of the wall-clock running time of a SHG calculation on bulk GaAs illustrates the scalability of the ArchNLO package. In Figure 2a, we plot the wall-clock time, the split time, and running time vs the number of nodes. The total wall-clock time equals the sum of the split time, which is the time for splitting the input data into groups of k-points, and the running time, which is the calculation time on each separated group of k-points for nonlinear susceptibilities. As shown in Figure 2a, the wall-clock time vs the number of nodes is nearly linear in the log scale, indicating that the ArchNLO package is in a nearly perfect scalability.

*Singularity problem:* It is known that there are a few special terms in Eq. 1~5, where the cancellation of energies in the denominator will lead to singularities [35,37,47,48]. These singularities occur under two conditions. The first is from the resonance term, $\omega_{mn} \approx \omega$ or $2\omega$, and this divergence can be avoided by adding a small positive damping factor $\eta$, making it be $\omega + i\eta$. Here $\eta$ is related to the bandwidth of Lorentzian resonance peaks. In our calculations for OHPs, we choose $\eta = 0.035\ eV$. The second is from the case, $\omega_{ln} - \omega_{ml} \approx 0$ at some special k-points. According to Ref. [37,47,48], such special k-points are rare, and their contributions are safe to be discarded. In our calculation, we set an energy tolerance $e_{tol}$ to decide which term to be discarded. The unit of $e_{tol}$ is $eV^2$ for linear optical calculations and $eV^4$ for second-order optical calculations. An appropriate choice of $e_{tol}$ is important for obtain smooth and converged SHG spectra. For example in Figure 2b, the choice of $e_{tol} = 10^{-1}$ discards too many terms and loses a few important features of the SHG spectrum of bulk GaAs. On the other hand, the choice of $e_{tol} = 10^{-20}$ keeps too many singularities. $e_{tol} = 10^{-4}$ is an appropriate choice of GaAs, showing a converged and smooth spectrum. For our calculation of OHPs, we choose $e_{tol} = 10^{-8}$ after testing different tolerant values.

## IV. Electronic Band Structures

Because all $CH_3NH_3MX_3$ structures share the same crystalline structure, their electronic band structures are essentially similar. We plot two representatives, i.e., $CH_3NH_3PbI_3$ and $CH_3NH_3SnI_3$, in Figure 3. The band gaps have been corrected by the scissor approximation based on HSE calculations. $CH_3NH_3MX_3$ exhibit a direct gap at the R point, which is at the corner of the first BZ as shown in Figure 1b. This result agrees with previous calculations [33].

There are a few important features in these band structures. First, SOC plays an important role to affect band gaps. In Figure 3, we can observe that SOC significantly reduces band gaps by a few hundred meV. Secondly, the electronic states around the band edges are mostly from the inorganic components of OHPs [27,33]. The projected density of states (PDOS) in Figure 3 shows that the bottom of the conduction band mainly originates from M atoms, while the top of the valence band is from halide atoms X as well as M atoms. These band-edge states almost do not hybridize with the $CH_3NH_3$ part. Therefore, the organic molecules mainly affect the crystal structure by breaking the inversion symmetry, rather than directly influencing optical responses around band edges.

Table 2 summarizes the band gaps of our studied OHP structures, and they are close to measurements. For example, the calculated band gap of $CH_3NH_3PbI_3$ is 1.64 eV which agrees with the measured value of 1.61 eV [49]. Table 2 also reveals that there is a monotonic correlation between the band gap and halide X atoms: the band gap increases from I to Br to Cl. Finally, it is worth mentioning that the above band-gap agreements between theory and measurements are not exactly *ab initio* results because they depend on the choice of the fraction of exchange in the hybrid functional, which is 0.45 in our calculations. Moreover, many extrinsic factors, such as octahedral tilting, substantially impact the measured band gaps as well. On the other hand, since we use the same hybrid functional for all OHPs, the evolution trend of band gaps shall be reliable.

## V. Linear Optical Response

The linear optical response of our studied $CH_3NH_3MX_3$ perovskites is presented in Figure 4. We focus on the optical absorption spectrum, which is the imaginary part of the linear dielectric function. By varying M and X elements, the optical absorption spectra can cover a wide spectrum from the near-infrared (NIR) frequency to the ultraviolet (UV) frequency. This is consistent with the corresponding changes of band gaps listed in Table 2. Moreover, following the trend of the halogen element changed from I, Br, to Cl, the intensity of the optical absorbance peaks is reduced. This is because the linear optical absorption is inversely proportional to the transition energy (band gap) according to Eq. 1. Finally, the anisotropy of optical absorption is enhanced when replacing the M atom from Pb, Sn to Ge, because of the stronger anisotropic structure shown in Table 1.

## VI. SHG

Our relaxation releases all symmetries of $CH_3NH_3MX_3$ perovskites. Therefore, all elements of the SHG tensor are expected to be nonzero. As an example, we present the absolute value of the elements of the SHG susceptibility tensor of $CH_3NH_3SnI_3$ in Figure 5. The subscripts 1, 2, and 3 denote the Cartesian coordinates *x*, *y*, and *z*. Because there are so many non-zero components, we group them by the magnitude of their zero-frequency values: $\chi^{(2)}_{311}$, $\chi^{(2)}_{122}$, $\chi^{(2)}_{111}$, $\chi^{(2)}_{133}$. and $\chi^{(2)}_{322}$ are plotted in Figure 5a as a group with the largest values; $\chi^{(2)}_{313}$, $\chi^{(2)}_{333}$ $\chi^{(2)}_{113}$, $\chi^{(2)}_{212}$, and $\chi^{(2)}_{223}$ are plotted in Figure 5b as a group with the second largest values; the rest components are small and are plotted in Figure 5c. The spatially averaged absolute value of SHG components is plotted in Figure 5d. We use a unit cell with a specific polarization direction of the organic unit in our calculations. However, in fabricated materials, because of finite temperature effect and different domains, the spatially averaged components may better reflect the realistic conditions and are close to measurements. Our further analysis show that, although the structural distortions and symmetry breaking are induced by the anisotropic organic molecule, SHG and corresponding optical transitions are mainly from transitions between electronic states of distorted cage $MX_3$ structures [50]. (see the details of discussion of anisotropic structural effects on SHG in Section III of the Supplementary Material [44])

Figure 6 shows that the averaged absolute values of components for $CH_3NH_3MX_3$. Meanwhile, component-resolved results are shown in Figure S5 (see the details in Section IV of the Supplementary Material [44]), where we focus on the group of $\chi^{(2)}_{311}$, $\chi^{(2)}_{122}$, $\chi^{(2)}_{111}$, $\chi^{(2)}_{133}$. and $\chi^{(2)}_{322}$, which have the most significant SHG response within the infrared frequency range. With the same M atom, the absolute value of SHG susceptibility $|\chi^{(2)}_{abc}|$ decreases along the trend of replacing the halide element from I to Cl. This is consistent with the trend of the increasing band gap because the SHG amplitude is inversely proportional to the square of transition energies, as shown in Eqs. 3-5. For instance, $CH_3NH_3SnI_3$, which owns the smallest band gap, has the largest SHG susceptibility: its $|\chi^{(2)}_{122}|$ and $|\chi^{(2)}_{133}|$ can reach about 500 pm/V at around 0.6 eV, and those along other directions, such as $|\chi^{(2)}_{111}|$, $|\chi^{(2)}_{311}|$, and $|\chi^{(2)}_{322}|$, also have broad peaks with amplitudes above 300 pm/V (see the details in Section IV of the Supplementary Material [44]). These values are

comparable to those of the widely used SHG materials, such as GaAs [51] and emerging two-dimensional transition-metal dichalcogenides [21]. Moreover, As shown in Figure S5, the largest SHG tenors are the *aaa* or *abb* components, indicating significant SHG for two incident photons along the same direction. This is beneficial for device implementations.

To analyze the spectra of SHG susceptibilities, it is helpful to compare them with the linear dielectric function. Take the (111) component of the SHG susceptibility as an example, which is plotted in Figure 7a. Meanwhile, the x-direction component of the linear (imaginary) dielectric function, $\varepsilon_x''(\omega)$, is plotted in a red-color curve in Figure 7b. Three differences can be observed. First, because of the second-order nature of SHG, the low-energy edge of SHG spectra is about half of that of the linear absorption spectra. Second, the peak positions are different between SHG and linear optical spectra. This is because of their different transition paths. Third, the amplitude of SHG spectra decreases faster with increasing energy than that of the linear response. This is because the SHG amplitude is inversely proportional to the square of the transition energy while that of the linear optical absorption is inversely proportional to the transition energy, according to Eq. 1~5.

On the other hand, Figure 7 reveals that there are strong correlations between SHG and the linear optical absorption. Following previous works [21], we replot the linear optical absorption with the double frequency, $\varepsilon_x''(2\omega)$ in Figure 7b (blue-color curves). This is an approximation to consider only two-phonon processes with identical energy, which reflects the double-photon resonance. Interestingly, as shown in Figure 7, within the frequency range confined by two vertical dashed lines, these two spectra agree with each other very well. This indicates that the main features (peaks) of SHG spectra, especially below the band gap ($E_g$), are dominated by double-resonance processes.

More specifically, we try to understand the large SHG susceptibility in $CH_3NH_3SnI_3$. For comparative purposes, we choose $CH_3NH_3PbI_3$ and $CH_3NH_3SnI_3$ and plot their imaginary parts (the blue-color curve) of the SHG susceptibilities in Figure 8. Meanwhile, the interband and intraband contributions [35,36,38] are plotted, respectively. The modulation contribution is much smaller and not listed here.

In Figure 8, the typical imaginary SHG spectra (black curve) of both $CH_3NH_3PbI_3$ and $CH_3NH_3SnI_3$ have two significant wave packages (I and II) with opposite phases, which are divided by the band gap approximately. We first focus on peak I, which is below the band gap. For the frequency range between $\frac{E_g}{2}$ and $E_g$, the signs of interband and intraband contributions are opposite to each other, resulting in a partial cancellation. Importantly, the intraband contribution (blue curve) of $CH_3NH_3SnI_3$ are significantly larger than its interband contribution (red curve), resulting in a higher peak I.

Then we focus on the higher-energy peak II, which is above the band gap ($E_g$). The interband and intraband contributions of $CH_3NH_3PbI_3$ still partially cancel each other. However, those of $CH_3NH_3SnI_3$ aggregate in both negative amplitudes for the frequency range between 1.5 and 2 eV, making its overall peak II higher than that of $CH_3NH_3PbI_3$. As a result, for both peak I and II regions, the SHG susceptibility of $CH_3NH_3SnI_3$ is larger than that of $CH_3NH_3PbI_3$.

It is worth mentioning that there is a mathematical reason for the opposite signs of interband and intraband contributions with the frequency range between $\frac{E_g}{2}$ and $E_g$. As explained in Figure 7, the double-resonance processes dominate the SHG within this frequency range. We can focus on those $2\omega$ items in the denominator of Eq. 3 and Eq. 4. Moreover, the first $2\omega$ term (with $\Delta_{mn}$) in Eq. 4 is smaller than the second one (without $\Delta_{mn}$) because $\Delta_{mn}$ is usually a small number [52]. Thus, we mainly compare the last term in Eq. 3 and Eq. 4. Most parts of these two terms have the same sign but the term in Eq. 4 has a negative front sign, making their overall signs opposite to each other. In fact, these characters have been widely observed in NLO spectra of many other solids [41,53,54].

## VII. LEO effect

The dielectric constants, zero-frequency SHG susceptibilities, the low-frequency LEO coefficients are summarized in Table 3. Among our studied structures, $CH_3NH_3SnI_3$ exhibits a large zero-frequency SHG component of $\chi^{(2)}_{311}$, around 101 pm/V, which greatly enhances its LEO coefficient. Using Eq. 6, the LEO coefficient $r_{311}$ of $CH_3NH_3SnI_3$ is found to be 3.7 pm/V, as shown in Table 3. This is comparable to those of GaAs [55] and CdTe [56].

To further illustrate the relation between the band gap, overall SHG spectrum, and the zero-frequency value of the SHG susceptibility that is directly related to the LEO coefficients, we construct a schematic picture in Figure 9 consisted of pairs of positive/negative Gaussian peaks with two peak values ($E_1, E_2$) and two linewidths ($\sigma_1^2$, $\sigma_2^2$) to mimic the peaks I and II in Figure 8. The real part of spectra is calculated through the Kramers-Kronig relation [21,35,36]. First, we focus on the band gap effect. In Figure 9a, we see that, from the orange curve to the green curve, which have the same linewidth $\sigma$, the two peaks of the imaginary spectrum move to higher energy because of a larger band gap of the green curve. Correspondingly, the zero-frequency value of the green curve of the real-part spectrum in Figure 9b decreases. This indicates that a smaller band gap contributes to a larger LEO effect.

Comparing the blue and orange curves, we see that, in Figure 9a, the lower-energy positive peak of the blue curve is significantly larger than its the higher-energy negative peak. This stronger lower-energy peak lead to a larger zero-frequency value of the real-part spectrum of the blue curve in Figure 9b.

This explains why $CH_3NH_3SnI_3$ has a larger zero-frequency value of the SHG susceptibility. As shown in Figure 8, the energies of the peaks I and II of $CH_3NH_3SnI_3$ are lower due to its smaller band gap. Moreover, the peaks I and II of $CH_3NH_3SnI_3$ are about the same scale while, for $CH_3NH_3PbI_3$, the negative peak II is stronger and broader than its positive peak I. Combining these two factors, the zero-frequency value of the SHG susceptibility as well as the LEO coefficient of $CH_3NH_3SnI_3$ are larger than those of $CH_3NH_3PbI_3$.

## VIII. Conclusion

In summary, we investigate the structural and electronic properties of $CH_3NH_3MX_3$ (M=Ge, Sn, Pb; X=halide) OHPs. Linear and second-order NLO properties, such as SHG and LEO effects, are explored by the ArchNLO package, which proves good parallelism and scalability on such large systems. Due to the lack of inversion symmetry, our studied OHPs exhibit SHG responses and LEO effect. Particularly, we find large SHG susceptibilities as well as LEO coefficients for $CH_3NH_3SnI_3$, which have application potentials in second-order NLO devices and LEO modulators. We further reveal that the large SHG susceptibilities and LEO coefficients of $CH_3NH_3SnI_3$ are due to its small band gap as well as large intraband double resonance contributions. This work may stimulate further experimental and theoretical investigations to achieve NLO responses in OHPs, which can be fabricated by low-cost solution-based approaches.

## Acknowledgments

W.S. and L.Y. are supported by the Air Force Office of Scientific Research (AFOSR) Grant No. FA9550-17-1-0304 and the National Science Foundation (NSF) CAREER Grant No. DMR1455346. G.Y.G. is supported by the Ministry of Science and Technology, National Center for Theoretical Sciences, and Academia Sinica Thematic Program (AS-TP-106-M07) of the Republic of China. This work used the computing resource from the Extreme Science and Engineering Discovery Environment (XSEDE), which is supported by NSF Grant No. ACI-1548562.

## AUTHOR INFORMATION

Corresponding Authors

*E-mail: gyguo@phys.ntu.edu.tw

*E-mail: lyang@physics.wustl.edu

**Figures and Tables**

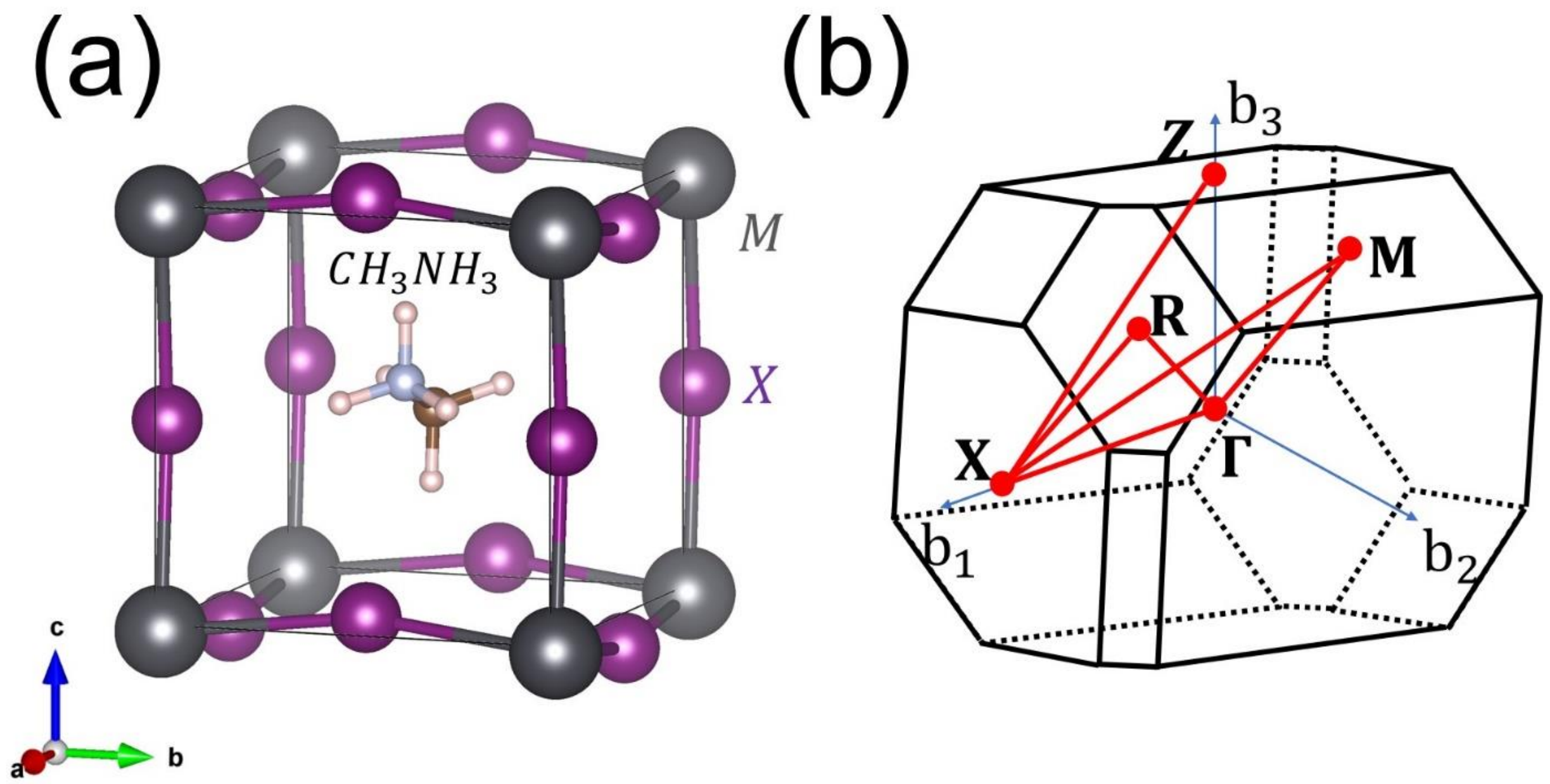

Figure 1. (a) Prototype atomic structure of our studied $CH_3NH_3MX_3$ (M=Ge, Sn, Pb; X= I, Br, Cl). (b) The first Brillouin Zone of the fully relaxed OHPs structure that is slightly distorted from the prototype cubic phase, making it essentially a triclinic phase. The structural details can be found in Table 1. High-symmetry points are marked.

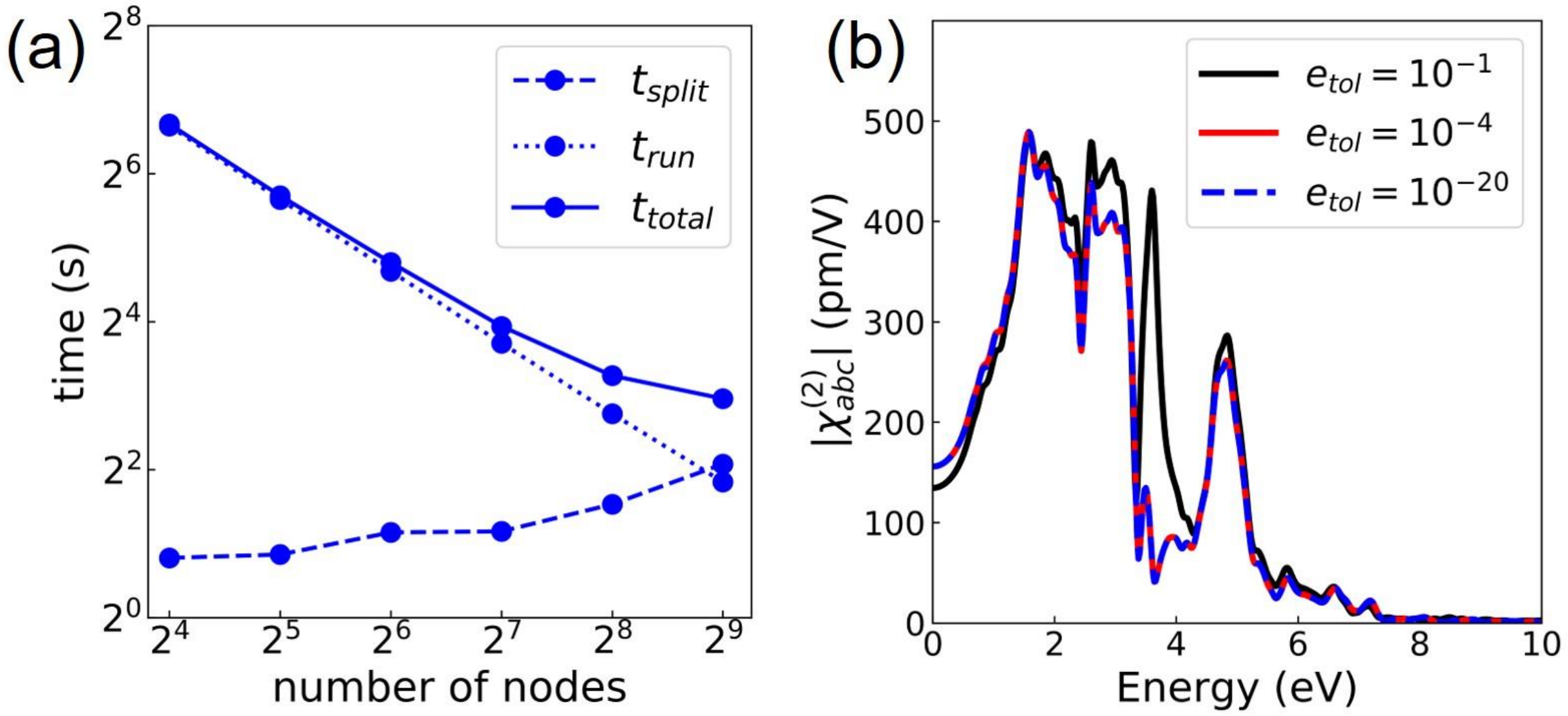

Figure 2. (a) Running time (in the $\log_2$ scale) of SHG calculations using ArchNLO under different number of nodes, in which $t_{total} = t_{run} + t_{split}$. The benchmark calculations are based on cubic GaAs with 40 total bands and an $8 \times 8 \times 8 = 512$ (non-reduced) k-point sampling. (b) Calculated SHG susceptibility with different energy tolerances (in a unit of $eV^2$) for cubic GaAs.

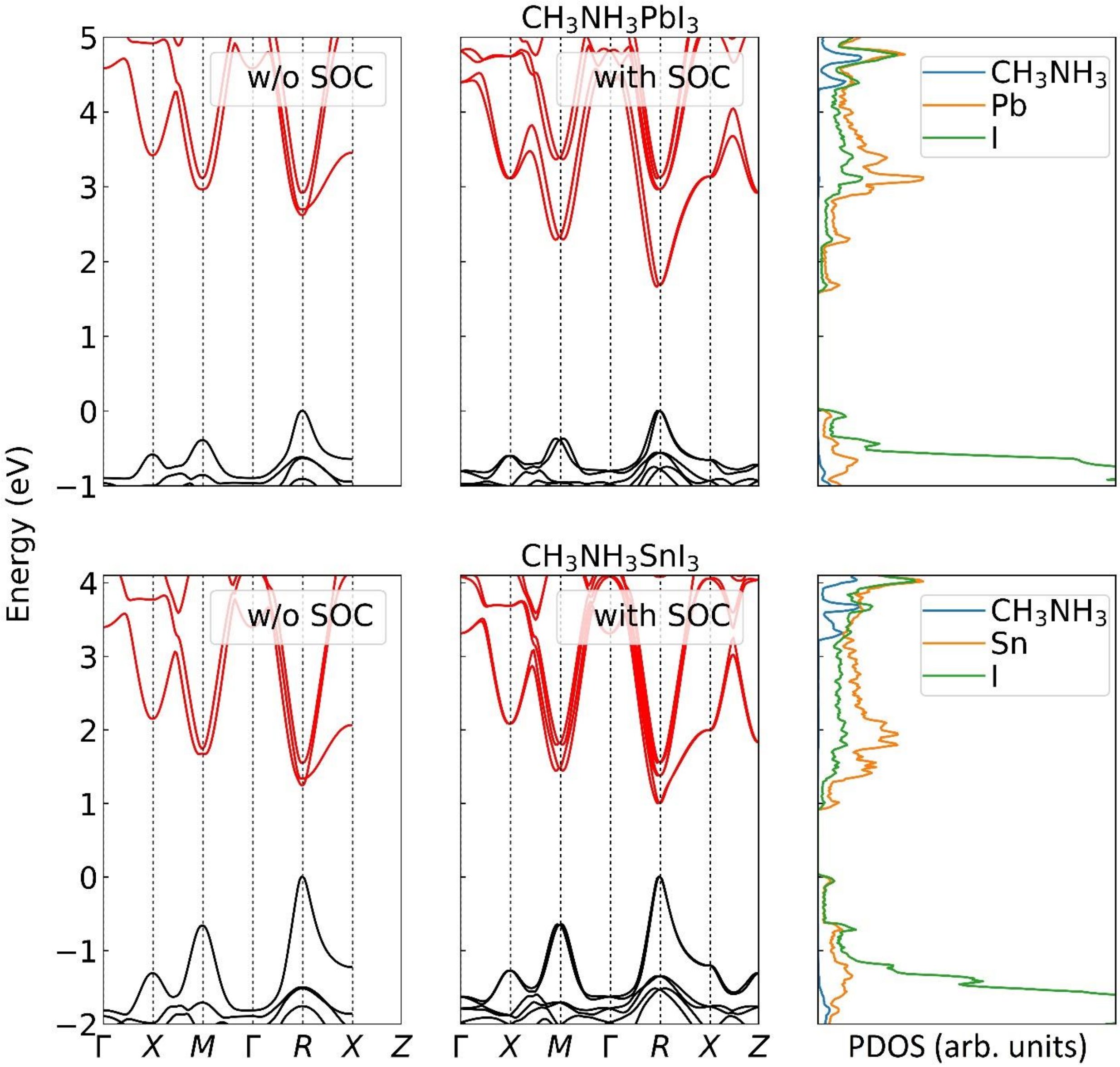

Figure 3. Band structures without SOC (left panels), with SOC (middle panel), and projected density of states (PDOS) with SOC (right panels) of $CH_3NH_3PbI_3$ and $CH_3NH_3SnI_3$. The band gap is corrected by HSE calculations. The top valence band is set to be 0 eV, and PDOS is in an arbitrary unit.

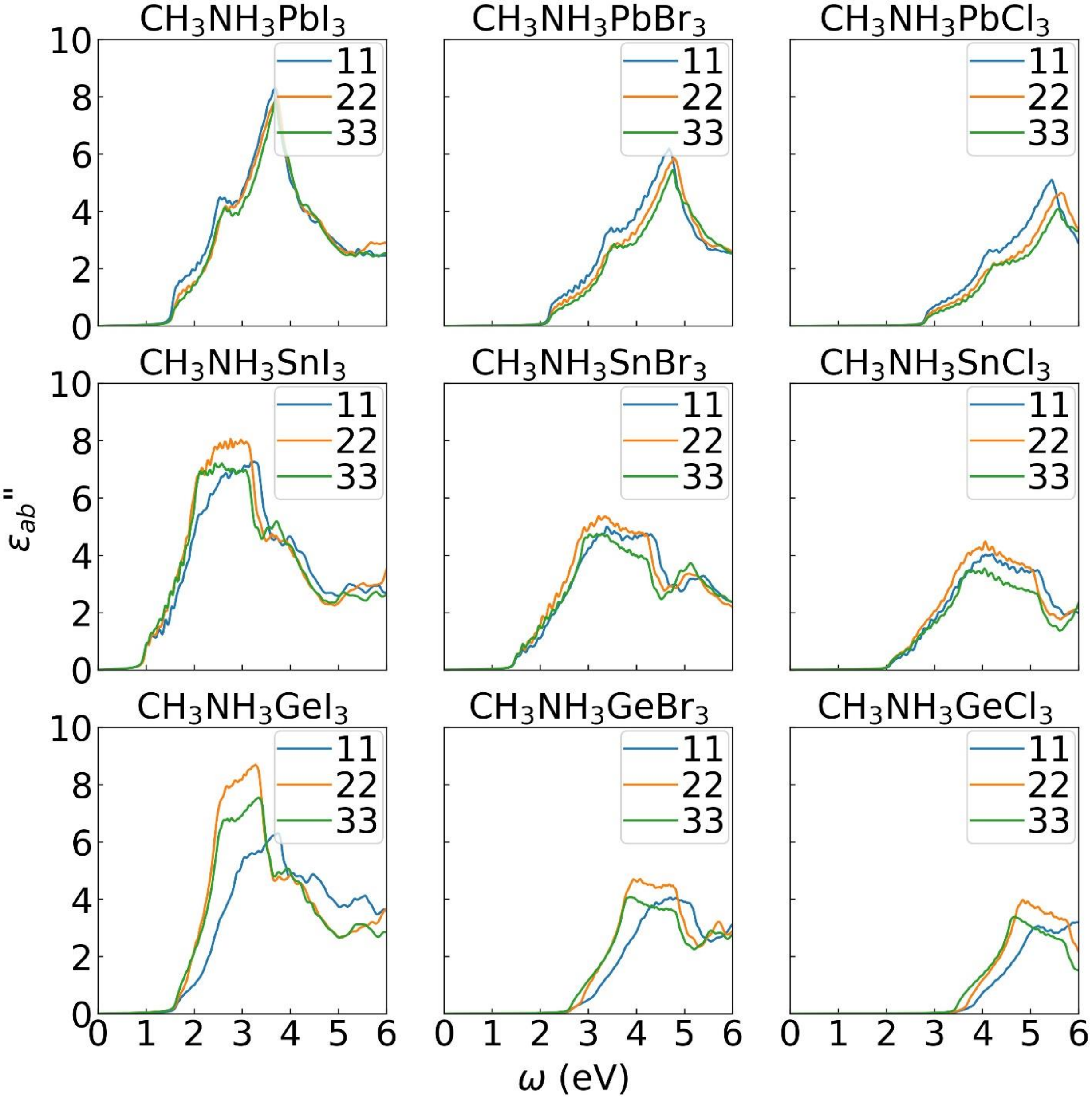

Figure 4. Diagonal components 11, 22, 33 of the imaginary linear dielectric function (dimensionless) of $CH_3NH_3MX_3$. The subscripts, 1, 2, and 3, denote the Cartesian coordinates x, y, and z.

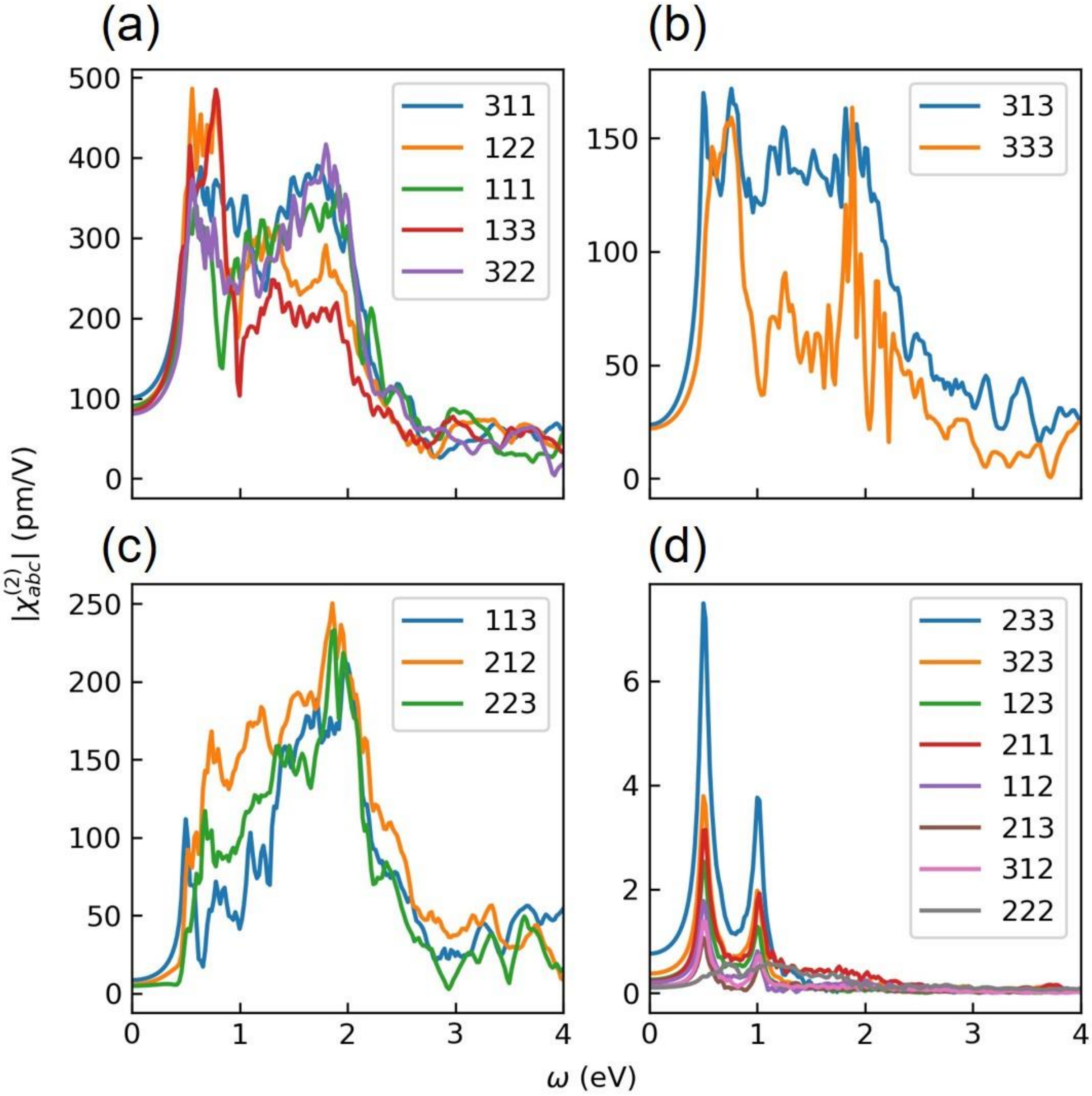

Figure 5. Absolute value of the SHG tensor ($\chi^{(2)}_{abc}$) of $CH_3NH_3SnI_3$. The subscripts, 1, 2, and 3, of the susceptibility denote the Cartesian coordinates x, y, and z. The four panels (a), (b), and (c) are grouped based on the rank of the magnitude of the zero-frequency value. The spatially averaged value of all components is plotted in (d).

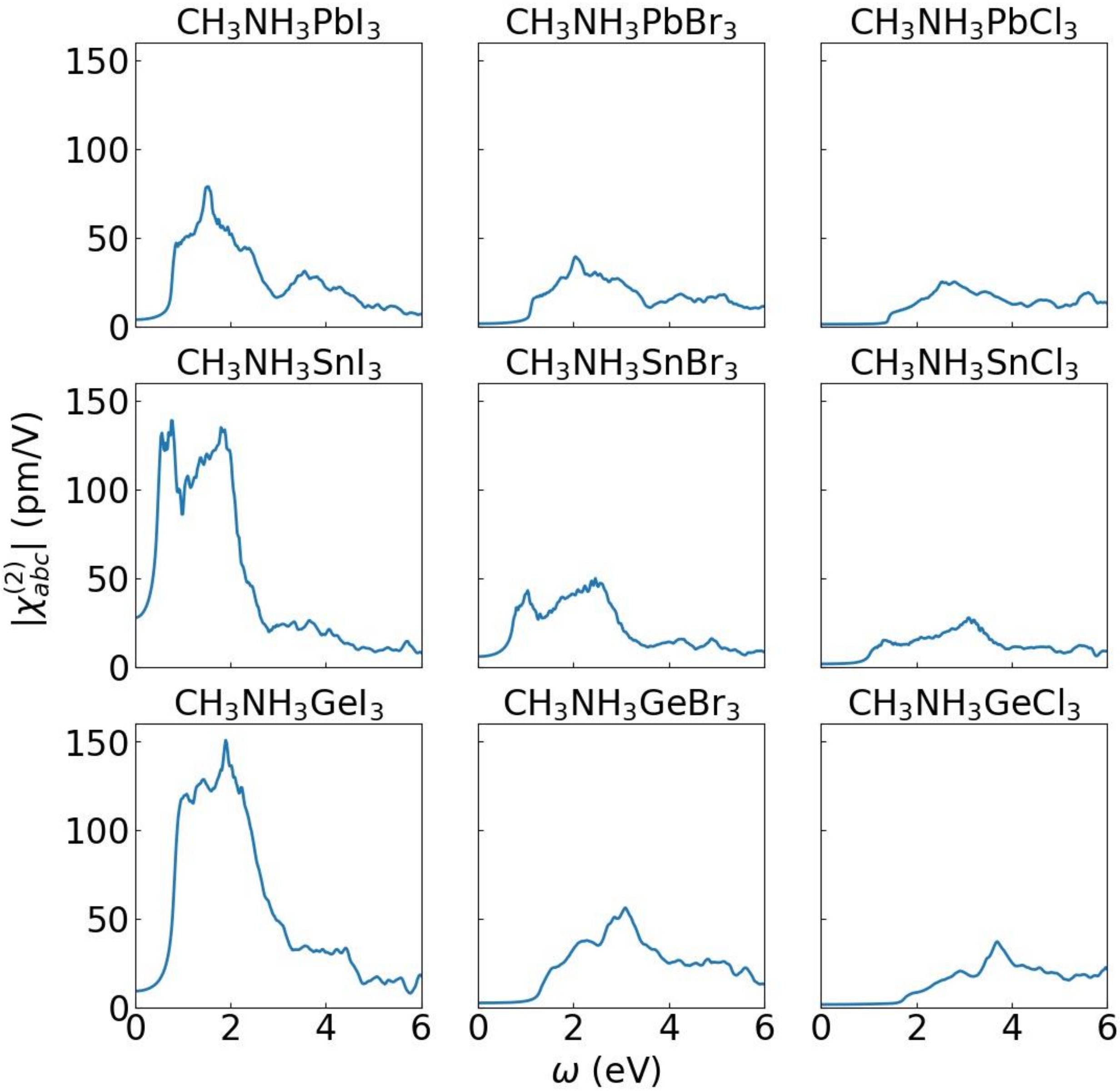

Figure 6. Spatially averaged absolute values of the SHG susceptibility of $CH_3NH_3MX_3$.

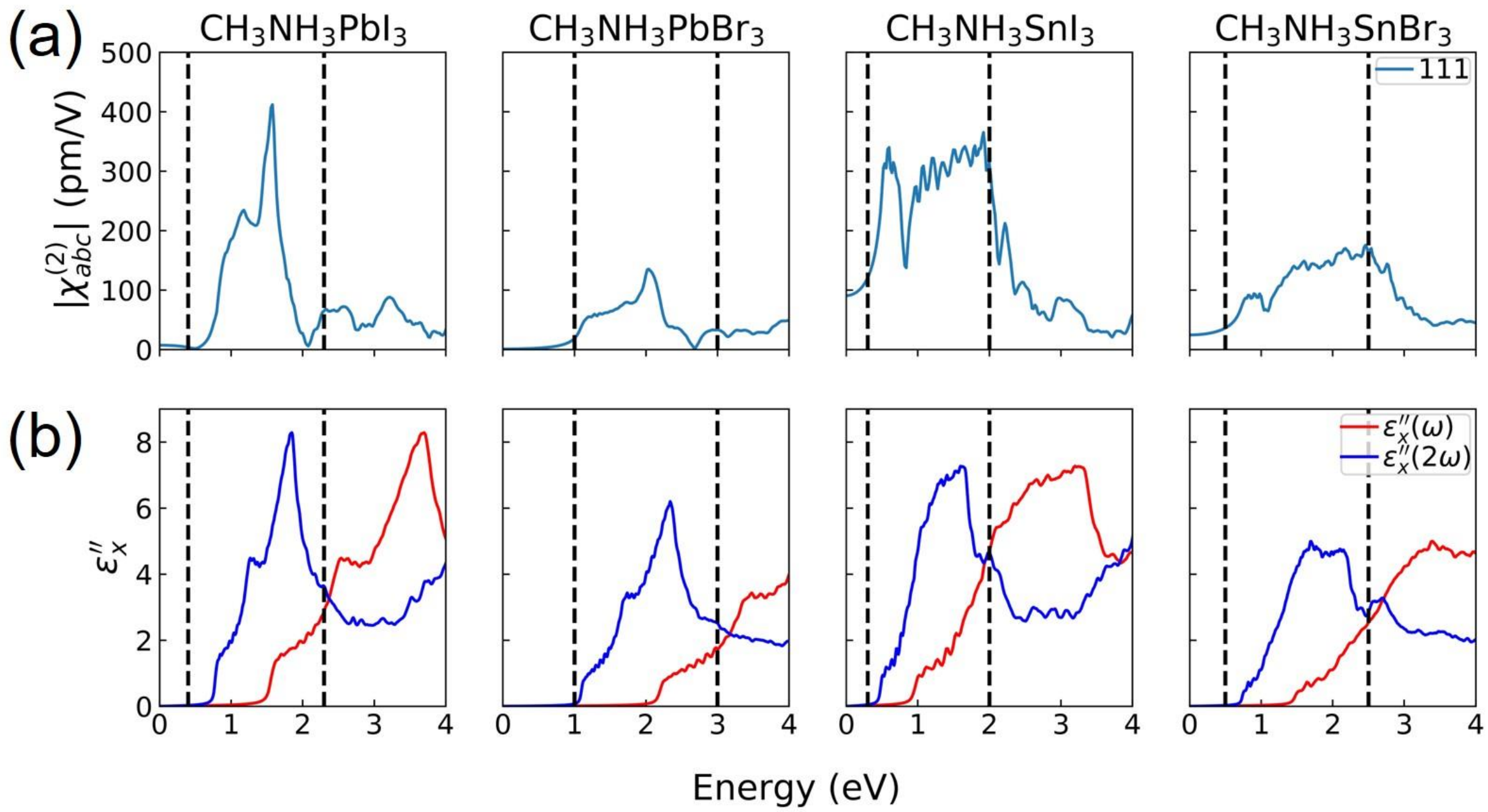

Figure 7. (a) Absolute values of ($|\chi^{(2)}_{111}|$) of the SHG tensor of $CH_3NH_3PbI_3$, $CH_3NH_3PbBr_3$, $CH_3NH_3SnI_3$, $CH_3NH_3SnBr_3$, and (b) the correspondent imaginary part of the dielectric function (dimensionless), $\varepsilon''_x(\omega)$, and the double-resonant one, $\varepsilon''_x(2\omega)$. The two vertical dashed lines mark the region for double resonance.

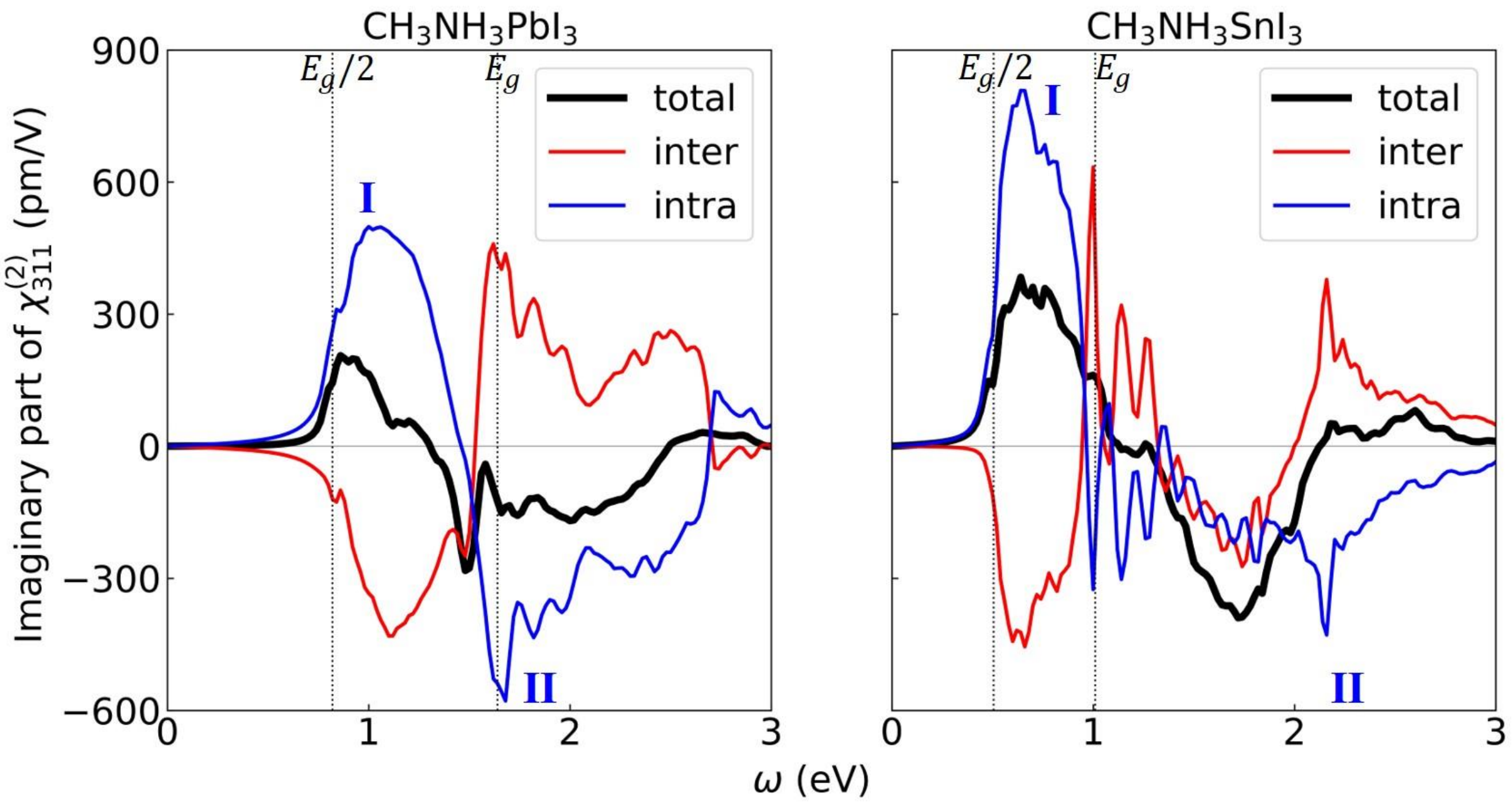

Figure 8. Imaginary part of the SHG susceptibility $\chi^{(2)}_{311}$ (total) and contributions from the interband (inter) and intraband (intra) terms of $CH_3NH_3PbI_3$ and $CH_3NH_3SnI_3$. Half band gap $E_g/2$ and band gap $E_g$ are marked with vertical black dot lines, respectively.

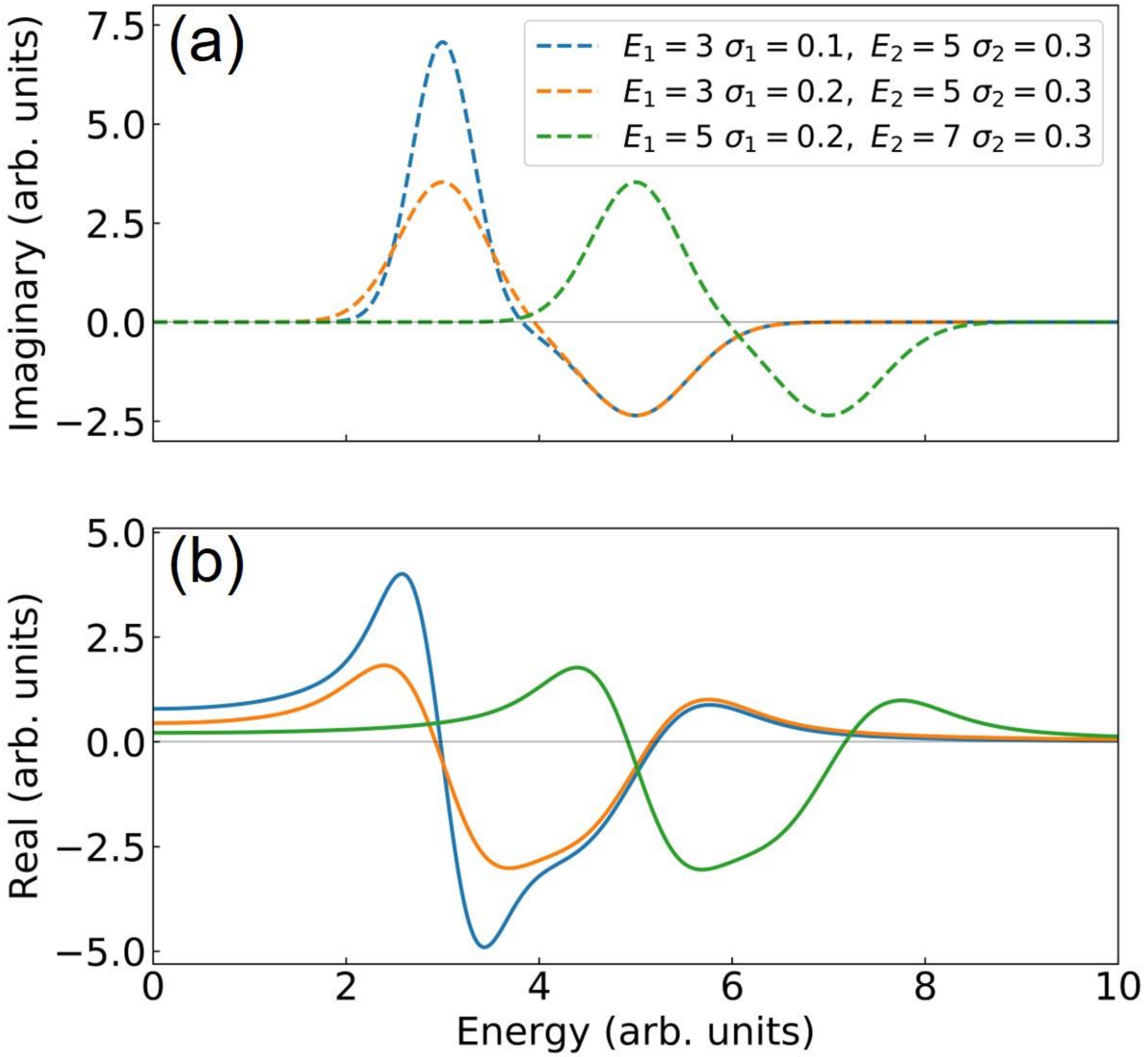

Figure 9. (a) Imaginary part of the SHG susceptibility with a pair of positive/negative Gaussian peaks and (b) real part of the SHG susceptibility obtained by the Kramers-Kronig relation. The SHG susceptibility is in an arbitrary unit.

**Table 1** Calculated lattice parameters of $CH_3NH_3MX_3$ OHPs.

| name | lattice parameters | | | | | |
|---|---|---|---|---|---|---|
| | a (Å) | b (Å) | c (Å) | $\alpha$ (°) | $\beta$ (°) | $\gamma$ (°) |
| $CH_3NH_3PbI_3$ | 6.31 | 6.23 | 6.38 | 90.0 | 89.2 | 90.0 |
| $CH_3NH_3PbBr_3$ | 5.92 | 5.86 | 6.02 | 90.0 | 88.9 | 90.0 |
| $CH_3NH_3PbCl_3$ | 5.65 | 5.59 | 5.76 | 90.0 | 89.0 | 90.0 |
| $CH_3NH_3SnI_3$ | 6.24 | 6.16 | 6.28 | 90.0 | 88.1 | 90.0 |
| $CH_3NH_3SnBr_3$ | 5.87 | 5.80 | 5.94 | 90.0 | 87.4 | 90.0 |
| $CH_3NH_3SnCl_3$ | 5.62 | 5.54 | 5.71 | 90.0 | 86.6 | 90.0 |
| $CH_3NH_3GeI_3$ | 6.11 | 5.98 | 6.02 | 89.5 | 85.5 | 88.8 |
| $CH_3NH_3GeBr_3$ | 5.80 | 5.70 | 5.79 | 89.5 | 83.5 | 88.6 |
| $CH_3NH_3GeCl_3$ | 5.54 | 5.44 | 5.57 | 90.7 | 82.4 | 89.5 |

**Table 2** HSE and DFT (in bracket) band gaps of $CH_3NH_3MX_3$ OHPs.

| Band gap (eV) | I | Br | Cl |
|---|---|---|---|
| Pb | 1.64 (0.72) | 2.24 (1.06) | 2.85 (1.49) |
| Sn | 1.01 (0.40) | 1.48 (0.60) | 2.10 (1.02) |
| Ge | 1.66 (0.89) | 2.63 (1.43) | 3.46 (2.01) |

**Table 3** Dielectric constants ($\varepsilon_x(0)$, $\varepsilon_y(0)$, and $\varepsilon_z(0)$), zero-frequency SHG susceptibility ($|\chi^{(2)}_{111}(0)|$, $|\chi^{(2)}_{122}(0)|$, $|\chi^{(2)}_{133}(0)|$, $|\chi^{(2)}_{311}(0)|$, $|\chi^{(2)}_{322}(0)|$, $|\chi^{(2)}_{average}(0)|$) and LEO coefficients, ($r_{111}, r_{122}, r_{133}, r_{311}, r_{322}, r_{average}$) of $CH_3NH_3MX_3$ OHPs.

| M | Pb | | | Sn | | | Ge | | |
|---|---|---|---|---|---|---|---|---|---|
| X | I | Br | Cl | I | Br | Cl | I | Br | Cl |
| $\varepsilon_x(0)$ | 5.9 | 4.5 | 4.0 | 7.2 | 5.2 | 4.4 | 5.9 | 4.1 | 3.7 |
| $\varepsilon_y(0)$ | 5.8 | 4.4 | 3.9 | 7.6 | 5.4 | 4.6 | 6.8 | 4.4 | 3.9 |
| $\varepsilon_z(0)$ | 5.6 | 4.2 | 3.7 | 7.3 | 5.1 | 4.2 | 6.4 | 4.1 | 3.7 |
| $\|\chi^{(2)}_{111}(0)\|$ $(pm/V)$ | 7.2 | 0.9 | 1.9 | 90.7 | 24.5 | 5.5 | 45.0 | 3.7 | 3.1 |
| $\|\chi^{(2)}_{122}(0)\|$ $(pm/V)$ | 6.8 | 3.0 | 0.3 | 90.9 | 17.7 | 1.3 | 9.1 | 4.4 | 3.0 |
| $\|\chi^{(2)}_{133}(0)\|$ $(pm/V)$ | 2.0 | 0.9 | 0.3 | 85.0 | 17.8 | 3.1 | 18.5 | 1.6 | 2.3 |
| $\|\chi^{(2)}_{311}(0)\|$ $(pm/V)$ | 6.1 | 0.6 | 1.9 | 101.2 | 19.5 | 0.9 | 14.2 | 2.8 | 2.7 |
| $\|\chi^{(2)}_{322}(0)\|$ $(pm/V)$ | 0.1 | 0.4 | 1.7 | 80.9 | 27.3 | 11.9 | 8.3 | 6.4 | 3.1 |
| $\|\chi^{(2)}_{average}(0)\|$ $(pm/V)$ | 4.0 | 1.6 | 1.3 | 27.7 | 6.6 | 1.9 | 9.3 | 2.6 | 1.7 |
| $r_{111}(0)$ $(pm/V)$ | 0.4 | 0.1 | 0.2 | 3.3 | 1.8 | 0.6 | 2.2 | 0.4 | 0.4 |
| $r_{122}(0)$ $(pm/V)$ | 0.4 | 0.3 | 0.0 | 3.3 | 1.3 | 0.1 | 0.5 | 0.5 | 0.4 |
| $r_{133}(0)$ $(pm/V)$ | 0.1 | 0.1 | 0.0 | 3.1 | 1.3 | 0.3 | 0.9 | 0.2 | 0.3 |
| $r_{311}(0)$ $(pm/V)$ | 0.4 | 0.1 | 0.3 | 3.7 | 1.4 | 0.1 | 0.7 | 0.3 | 0.4 |
| $r_{322}(0)$ $(pm/V)$ | 0.0 | 0.0 | 0.2 | 3.0 | 2.0 | 1.2 | 0.4 | 0.7 | 0.5 |
| $r_{average}(0)$ $(pm/V)$ | 0.2 | 0.2 | 0.2 | 1.0 | 0.5 | 0.2 | 0.5 | 0.3 | 0.2 |